\documentclass{article}

\usepackage[x11names, rgb]{xcolor}
\usepackage[utf8]{inputenc}
\usepackage{tikz}
\usetikzlibrary{snakes,arrows,shapes}
\usepackage{mathtools}
\usepackage{amssymb}
\usepackage{amsmath}
\usepackage{textcomp}
\usepackage{multirow}
\usepackage{algpseudocode,algorithm,algorithmicx}
\usepackage{wrapfig}
\usepackage{url}
\DeclareGraphicsExtensions{.pdf,.jpeg,.png}
\usepackage{pgf}
\usepackage{amsthm}
\newtheorem{definition}{Definition}
\newtheorem{theorem}{Theorem}
\newtheorem{example}{Example}

\newcommand{\U}{\ensuremath{\,\mathcal{U}\,}}

\newcommand{\X}{\ensuremath{\mathcal{X}\,}}
\newcommand{\F}{\ensuremath{\Diamond\,}}
\newcommand{\G}{\ensuremath{\Box\,}}

\newcommand{\words}{\ensuremath{\mbox{\textit{Words}}}}

\newcommand{\NBA}{B\"uchi Automa}

\newcommand{\spot}{\textsc{spot}}

\newcommand{\smv}{\textsc{smv}}

\pgfdeclareimage[interpolate=true,width=300pt]{nba}{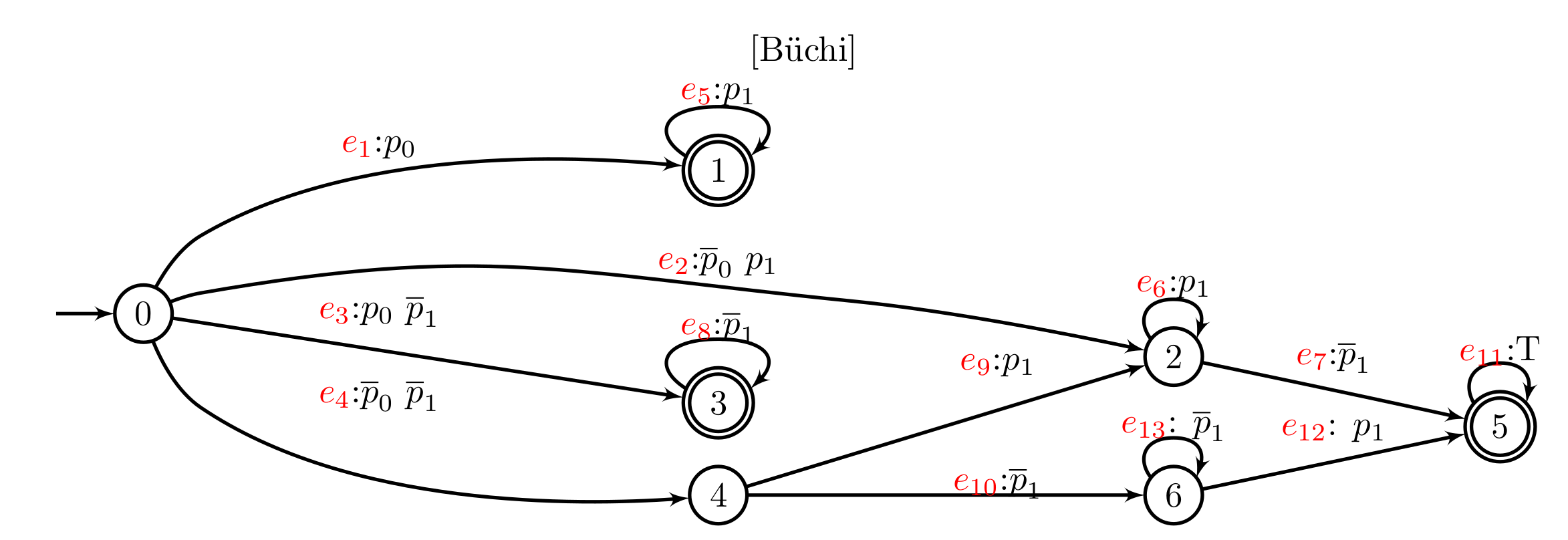}
\pgfdeclareimage[interpolate=true,width=200pt]{mealy}{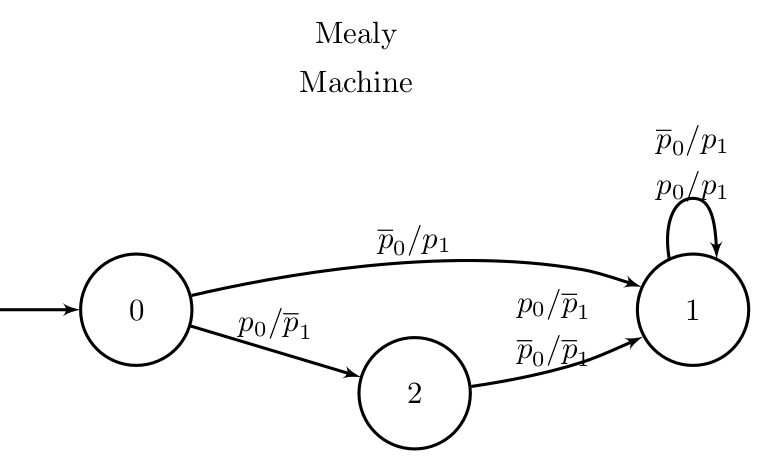}
\pgfdeclareimage[interpolate=true,width=150pt]{monitor}{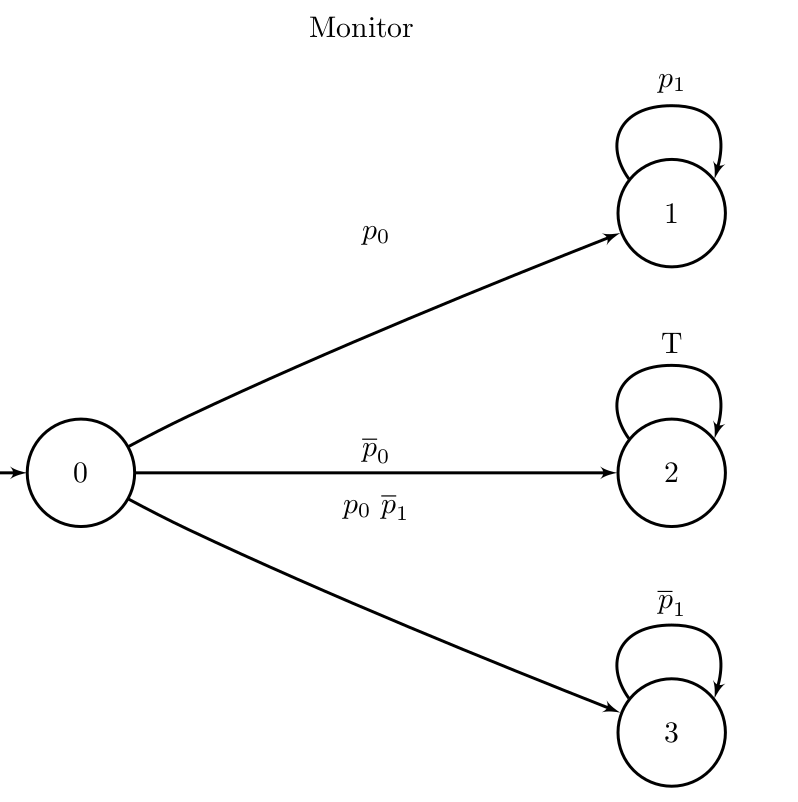}

\title{Automated Requirements-Based Testing\\ of Black-Box Reactive Systems}

\author{Massimo Narizzano$^1$, Luca Pulina$^2$,\\
	Armando Tacchella$^1$ and Simone Vuotto$^{1,2}$\\
	$^1$DIBRIS, University of Genoa, \\Viale Causa 13, 16145
	Genoa, Italy\\ \texttt{massimo.narizzano@unige.it,
		armando.tacchella@unige.it}\\ 
	$^2$Chemistry and Pharmacy Dept., University of Sassari, \\Via Vienna 2, 07100
	Sassari, Italy\\ \texttt{lpulina@uniss.it, svuotto@uniss.it}
}

\date{ \ }

\begin{document}
\maketitle

\begin{abstract}

We present a new approach to conformance
testing of black-box reactive systems.
We consider system specifications written as linear temporal logic
formulas to generate tests as sequences of input/output pairs:
inputs are extracted from the B\"uchi automata corresponding to the
specifications, and outputs are obtained by feeding the inputs
to the systems. Conformance is checked by comparing 
input/output sequences with automata traces to detect
violations of the specifications. 
We consider several criteria for extracting tests and for stopping
generation, and we compare them experimentally using both 
indicators of coverage and error-detection.
The results show that our methodology can generate test suites
with good system coverage and error-detection capability. 

\end{abstract}

\section{Introduction}
\label{sec:intro}
We are concerned with the 
problem of checking whether a reactive system --- which we can
execute, but for which we have no internal representation --- conforms
to a set of requirements provided as 
temporal logic formulas. This problem arises in a variety of 
contexts, e.g., when a system is developed by integrating
commercial off-the-shelf (COTS) componenents~\cite{li2009development}.
In these scenarios, 
techniques such as model checking~\cite{baier2008principles} or
(white-box) model-based testing~\cite{utt06} are ruled out. 
Also, classical black-box techniques like random testing, equivalence
partitioning or boundary  analysis~\cite{burnstein2006practical}
either do not take into account the specification or require manual
effort to assemble meaningful test suites. Techniques aimed
at automated test generation for black-box reactive systems relying on 
formal models of the specifications have been explored --- see,
e.g.,~\cite{krichen2004,BarbotBD19,koch98,schmitt00,jard05} --- and
they seem more promising than classical techniques when both
efficiency of test generation and effectiveness in covering the
specification are considered.   
Runtime verification~\cite{bauer2011} techniques 
can be seen as a
form of \textsl{oracle-based testing}~\cite{bernot1991software}:
each test is executed on the system implementation and the test oracle, i.e.,
the monitor in runtime verification jargon, observes the system
and checks whether its executions are behaviors allowed by the
specification or not. 
Following this stream of research, a technique based on the use of 
monitors as test oracles is proposed in~\cite{arcaini2013online}.
Their approach can test for safety properties (``something bad will
never happen''), but it does not deal with liveness properties
(``something good will happen infinitely often''). While
liveness properties are not amenable to monitoring on finite executions,
their proper subclass of co-safety properties (``something good
will happen'') consists of formulas that can be monitored on finite
traces and that we wish to consider when testing a system for
conformance.   

Our approach is inspired by~\cite{arcaini2013online}, but aims to
deal with a more general class of properties. 
Our methodology is based on a visit of the B\"uchi
automaton corresponding to the requirements. The visit 
starts from the initial state of the automaton and generates a
sequence of input values with which the black-box system is fed to
obtain a corresponding sequence of output values.  
We check such input/output sequence against the automaton, i.e., we
check whether there exists at least one state in the automaton that
can be reached along the sequence. 
If there is no such state, then the system is not conformant
to the requirements and the sequence provides a counterexample.
Otherwise, we can continue the generation of the sequence by iterating
the above steps until either $(i)$ 
an acceptance state of the automaton is reached with a sequence
of length at least $k_{min}$ or $(ii)$ an
acceptance state cannot be reached with a sequence of length at most
$k_{max}$, where $k_{min}$ and $k_{max}$ are two parameters such that
$k_{min} < k_{max}$. 
Multiple tests can be obtained by iterating this procedure
until all the reachable transitions have been visited at least once.

We evaluate our approach in three different experimental settings.
In the first one we consider benchmarks taken from
the LTL Track of the 2018 edition of the Reactive Synthesis
Competition (SYNTCOMP 2018)\footnote{\url{http://www.syntcomp.org/}} 
and we compare our approach with the one described
in~\cite{arcaini2013online}. 
In the second setting we use the Adaptive Cruise Control (ACC)
prototype implemented in~\cite{aniculaesei2018automated} and we
compare the tests generated by our approach with those
generated with a model-based generation strategy. 
In the third setting we test the model of a robotic arm
controller in order to evaluate our approach on a large set of
requirements coming from an industry-grade prototype.
In the two former settings we use a mix of
fault-injection~\cite{hsueh1997fault} and mutation
analysis~\cite{andrews2006using} in order to compare  
different approaches. In the third setting we inject faults manually.
The results we obtained with our experiments show that
our approach can outperform the one in~\cite{arcaini2013online}
by finding more induced faults. Furthermore, 
generating tests based on the specification
can be as effective as approaches based on the system model,
discovering almost the same number of faults.
Finally, our approach can be effective in finding faults in
small-to-medium sized industry-grade systems.

The rest of the paper is structured as follows. In
Section~\ref{sec:background} we present some basic notation and
definitions. In Section~\ref{sec:algo} we describe our framework for
test case generation of black-box system.
Finally, in Section~\ref{sec:exp} we show experimental results and we
conclude the paper in Section~\ref{sec:concl} with some remarks and an
agenda for future work.

\section{Preliminaries and Related Work}
\label{sec:background}
In this Section we recall the basic concepts used trough the
paper. First, we present some basic definitions, followed by syntax
and semantics of Linear Temporal Logic (LTL). Then we provide a short
introduction to $\omega$-regular grammars and languages and we
conclude the section by presenting related work.

\subsection{Non Deterministic B\"uchi Automa}

\begin{definition} [Non Deterministic B\"uchi Automata] A non deterministic B\"uchi
  Automata (NBA) $\mathcal{A}$ is a tuple $\mathcal{A}$ =
  (Q, $\Sigma$, $\delta$, $q_0$, $F$) where:
  \begin{itemize}
  \item Q is a finite set of states,
  \item $\Sigma$ is an alphabet,
  \item $\delta$ : $Q \times \Sigma$ $\rightarrow$ $2^Q$ is a transition function
  \item $q_0 \in Q$ is the initial state
  \item $F\subseteq Q$ is a set of accept states, called
    acceptance set.
  \end{itemize}
\end{definition}
Let $\Sigma^\omega$ denote the set of all infinite words over the
alphabet $\Sigma$. 

\begin{definition} [Run] A run for an infinite word $\sigma$ =
  $A_0A_1A_2... \in \Sigma^\omega$ denotes an infinite sequence
  $\varrho$= $q_0q_1q_2...$ of states in $\mathcal{A}$ such that
  $q_0 \in Q_0$ and $q_{i+1}=\delta(q_i,A_i)$ for i $\geq$ 0, and
  $\forall A_i$, $A_i\in\Sigma$.
  
\end{definition}
Notice that each run $\varrho$ in a NBA \textsl{induces} a
corresponding word  $\sigma \in \Sigma^{\omega}$.

\begin{definition} [Accepting run] A run $\varrho$ is accepting if
  there exist $q_i \in F$ such that $q_i$ occurs infinitely many times
   in $\varrho$.
\end{definition}



\begin{figure}[t!]
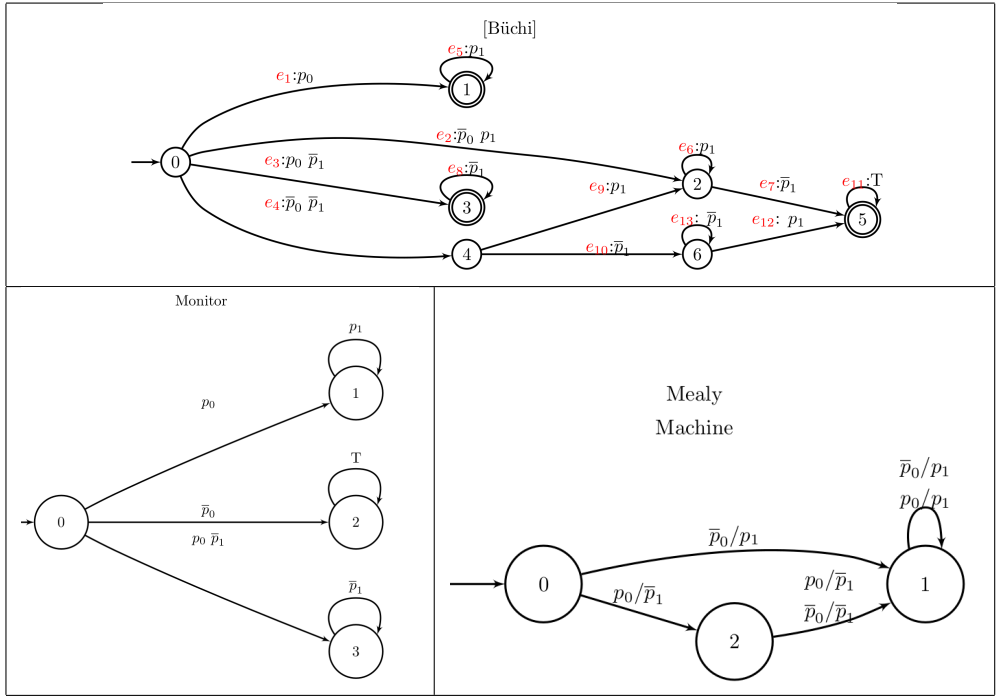

\begin{center}
 \begin{tabular}{|l|l|} \hline
\multicolumn{2}{|c|}{\pgfuseimage{nba}} \\\hline
\pgfuseimage{monitor} & \pgfuseimage{mealy} \\\hline
\end{tabular}
\end{center}
  \caption{\label{fig:nbamonitormealy} A state-based B\"uchi automaton (top), the corresponding monitor (bottom-left) and a Mealy machine (bottom-right). We write $\overline{p}_i$ to denote $\neg p_i$.}
\end{figure}
Figure~\ref{fig:nbamonitormealy} (top), shows a NBA where
$Q$=$\{0,1,2,3,4,5,6\}$, $\Sigma$=$2^{AP}$, $AP$=$\{p_0,p_1\}$,
$q_0$=$0$, and $F$=$\{1,3,5\}$. Throughout the paper we make use
of propositional logic formulae as a shorthand notation for the
transitions of NBAs. For instance, a label $a \vee b$ on an edge from
a state $q$ to a state $p$, represents three transitions from $q$ to $p$: one 
for the symbol $\{a\}$, one for the symbol $\{b\}$, and one for the
symbol $\{a,b\}$.

\subsection{LTL syntax and semantics}\label{sec:LTLdef}

Linear temporal logic (LTL)~\cite{pnueli1977} formulae consist of
atomic propositions, Boolean operators, and temporal operators. The
syntax of a LTL formula $\phi$ is given as follows:
\begin{center}
  $\phi =$ $\top$ $\mid$ $\bot$ $\mid$ $a$ $\mid$ $\neg \phi_1$
  $\mid$ $\phi_1\vee\phi_2$ $\mid$ $\X \phi_1$ $\mid$ $\phi_1
  \U \phi_2$ $\mid$ $( \phi )$ 
\end{center}
where $a \in AP$, $\phi, \phi_1, \phi_2$ are LTL
formulae,  $\mathcal{X}$ is the ``next'' operator and $\mathcal{U}$ is
the ``until'' operator. In the following, unless specified otherwise
using parentheses, unary operators have higher precedence than binary
operators. We also write $\overline{\phi}$ to denote $\neg \phi$.

Informally, the semantics of an LTL formula $\phi$ can be defined over
the language that contains all infinite words over the alphabet
2$^{AP}$. More precisely: 

\begin{definition}[Set of words over $2^{AP}$] 
  Given a set of atomic propositions $AP$, $(2^{AP})^\omega$ denotes
  the set of $words$ that arise from the infinite concatenation of
  symbols from the alphabet $(2^{AP})$.  Each word is defined as
  $\sigma=$ $A_0$$A_1$$A_2\dots$ $\in$ $(2^{AP})^\omega$, where each
  $A_i$ is a set over $AP$, i.e. $A_i\in 2^{AP}$.
\end{definition}

In the following, for $\sigma=$ $A_0$$A_1$$A_2\dots$ $\in$
$(2^{AP})^\omega$, $\sigma[j\dots]=$ $A_j$$A_{j+1}\dots$ $\in$
$(2^{AP})^\omega$ is the suffix of $\sigma$ starting in the $(j+1)$st
symbol $A_{j}$. 


\begin{definition}[LTL semantics over words]
  Let $\phi$ be an LTL formula over the set $AP$ and 
  let $\sigma$=$A_0A_1A_2\dots$ be an infinite word over ($2^{AP}$).
  We define the relation ``$\models$'' between $\sigma$ and $\phi$ as
  as the smallest relation with the following properties:
  \begin{enumerate}
  \item $\sigma$ $\models$ $true$
  \item $\sigma$ $\models$ a   iff a $\in$ $A_0$
  \item $\sigma$ $\models$ $\phi_1 \wedge \phi_2$  iff $\sigma$ $\models$ $\phi_1$ and $ \sigma  \models \phi_2$ 
  \item $\sigma$ $\models$ $\neg$ $\phi$   iff $ \sigma  \not\models \phi$
  \item $\sigma$ $\models$ $\X$ $\phi$   iff $ \sigma[1...]=A_1A_2A_3 \models \phi$    
  \item $\sigma$ $\models$ $\phi_1 \U$ $\phi_2$ iff $ \exists j \geq
    0$ such that $\sigma[j...]=A_jA_{j+1}... \models \phi_2$
    and $\sigma[i...]\models \phi_1$ $\forall 0 \leq i < j$
  \end{enumerate}
\end{definition}

We consider other Boolean connectives like ``$\wedge$'' and 
``$\rightarrow$'' with the usual meaning, while we introduce $\F \phi$
(``eventually'') to denote $\top \U \phi$ and $\G \phi$ (``always'')
to denote $\neg \F \neg \phi$. 

\begin{definition}[Accepted Words for a LTL formula]
We also define the set of accepted Words of a LTL formula $\phi$ as
the set containing all the infininte word $\sigma$ over $2^{AP}$ that
satisfy the property $\phi$, i.e.

\centerline{Words($\phi$) = \{$\sigma$ $\in$ $2^{AP}$ $|$ $\sigma$ $\models$ $\phi$\}}

\end{definition}

\begin{theorem}[Constructing an NBAs for an LTL formula~\cite{baier2008principles}]
For any LTL formula $\phi$ (over AP) there exists an NBA $\mathcal{A}_{\phi}$ with
Words($\phi$) = $\mathcal{L}_{\omega}(\mathcal{A}_{\phi})$.
\end{theorem}
\begin{example}
Figure~\ref{fig:nbamonitormealy} (top), shows a NBA obtained from
the formula\\
\centerline{$p_0$ $\leftrightarrow$ (\X \G $p_1$ $\vee$ $\overline{\Diamond p_1}$)}
where $AP = \{p_0,p_1\}$. The NBA is obtained using \spot{}~\cite{spot}.~\footnote{Using the command line \textsl{ltl2tgba} -B -f
``p0 $<$$-$$>$ (X G p1 $|$ ! F p1)''}
\end{example}

\begin{definition}[Mealy machine]
A Mealy machine is a tuple $\mathcal{M}$ = (S, $s_0$, $I$, $O$,
$\delta$) where:   
  \begin{itemize}
  \item S is a finite set of states,
  \item $s_0 \in S$ is the start state  
  \item $I$ is a set of symbols called input alphabet,
  \item $O$ is a set of symbols called output alphabet,
  \item $\delta$ : $S \times I$ $\rightarrow$ $S \times O$ is a
  transition function mapping pairs of states and input symbols
  to the corresponding pairs of states and output symbols
  \end{itemize}  
\end{definition}
In other words, a Mealy machine is a finite-state machine whose output
values are determined by its current state and the current inputs. 
\begin{example}
Figure~\ref{fig:nbamonitormealy} (bottom-right) shows a Mealy machine
obtained by using STRIX~\cite{strix} on the formula
$$
p_0 \leftrightarrow (\X \G p_1 \vee \overline{\Diamond p_1})
$$
where  S =$\{0,1,2\}$, $s_0$ = 0,  I =$\{p_0\}$ and   O =$\{p_1\}$.
\end{example}

\subsection{Monitor}
A monitor is an automaton supposed to follow the execution
of a system and move accordingly. An error is detected when 
the monitor cannot move, i.e., the system has performed some action,
or reached some state that it was not meant to be.

\begin{definition} [Monitor] A monitor $\mathcal{M}$ is a tuple $\mathcal{M}$ =
  (Q, $\Sigma$, $\delta$, $q_0$) where:
  \begin{itemize}
  \item Q is a finite set of states,
  \item $\Sigma$ is an alphabet,
  \item $\delta$ : $Q \times \Sigma$ $\rightarrow$ $2^Q$ is a transition function
  \item $q_0 \in Q$ is the initial state
  \end{itemize}
\end{definition}

\begin{example}
Figure~\ref{fig:nbamonitormealy} (bottom-left) shows a monitor obtained
using \spot{}~\cite{spot}~\footnote{Fired with command line \textsl{ltl2tgba} -MD -f ``p0 $<$$-$$>$ (X G p1 $|$ ! F p1)''.} for the formula
$$
p_0 \leftrightarrow (\X \G p_1 \vee \overline{\Diamond p_1})
$$
where $Q$=$\{0,1,2,3\}$, $\Sigma$=$2^{AP}$, $AP$=$\{p_0,p_1\}$, and $q_0$=$0$.
\end{example}

\subsection{Related Work}

The research most closely related to ours is presented
in~\cite{arcaini2013online} where the authors describe a methodology
for online testing of Java classes. Their key technique is to exploit a
monitor derived from LTL specifications to check conformance of the
system to stated requirements, with a focus on safety properties. In
order to compare this mehodology with our approach, we 
reimplemented the idea presented in~\cite{arcaini2013online}, 
making it applicable to any black-box system and not just Java
classes. Another work related to ours is presented
in~\cite{krichen2004} where the authors describe a methodology for
specification based testing of black-box systems. They assume that the
specification of the system is given as a non-blocking input/output
timed automaton, and the system itself --- whose model need not to be
known --- is also a timed automaton. The two main differences between
their methodology and ours are $(i)$ the capability of dealing with 
real-time requirements and $(ii)$  the form of the specification: ours
is ``declarative'', in the form of a set of LTL requirements, whereas
theirs is ``operational'' in the form of an automaton. We thus incur
into one additional step, i.e., extracting an automaton from the
requirements, after which the two methodologies proceed in a similar
way. However, given the different form and expressivity of the
requirements, a direct comparison is not easily feasible, and might be
even misleading. More recently in~\cite{BarbotBD19}, another approach
based on timed automata to specify input signals constraints has been
proposed. Also this approach bears some similarity with ours and with
that of~\cite{krichen2004}, but in our opinion it is not directly
comparable, at least in the settings that we consider for our
experimental analisys.

Other research which is closely related to ours appears in a series of
papers~\cite{tan2004,zeng2016test,zeng2015test} where the authors
present a test-case generation methodology that $(i)$ translates LTL 
requirements into Generalized B\"uchi Automata, $(ii)$ builds trap
properties from them --- using different criteria --- and $(iii)$
performs model checking of negated trap properties against the system
model in order to extract test cases.
The main difference with our work is that such methodology relies on
a model of the system under testing, a model that must be
verified against the system specification. Failing to do so, may
generate conflicting tests, i.e., a test which fulfills a requirement,
and violates another. To the extent of our knowledge there is no other
recent work on formally-grounded methods for requirement based
testing, while there is some not-so-recent work mentioning
conformance testing to specification, such as, for
example~\cite{koch98,schmitt00,jard05}. However, in these works
specifications are mostly ``operational'' in the form, e.g., of finite
state machines and thus a direct comparison with our methodology is
not possible.

%
%

\section{Automatic Test Case Generation from LTL specification}
\label{sec:algo}

\begin{figure}[t]
	\includegraphics[width=\textwidth]{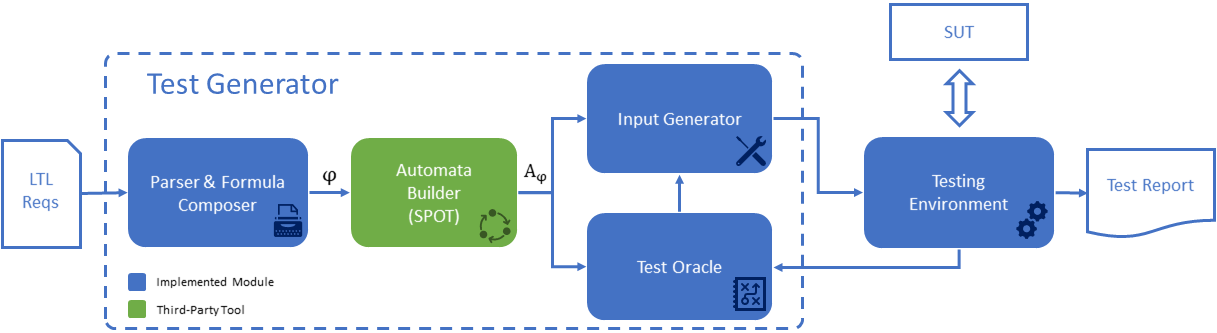}
	\caption{The main workflow of our approach.}
	\label{fig:framework}
\end{figure}

In order to test black-box systems, our approach adopts the workflow
presented in Figure \ref{fig:framework}.  
We assume that the specification is composed of a list of LTL
formulas, the declaration of the set $I$ of input propositions, and
the set $O$ of output propositions 
such that $I \cup O = AP$ and $I \cap O = \emptyset$.  The
``Test Generator'' pipeline in Figure~\ref{fig:framework} has the goal to
produce a set of valid tests to execute on the system under test (SUT).
The pipeline comprises four components:
\begin{itemize}
\item \textsf{Parser} reads the input specification, creates
  the intermediate data structures and builds the conjunction of
  requirements.
\item \textsf{Automata Builder} builds a B\"uchi or
  equivalent automaton representation of the input specification.
\item \textsf{Input Generator} chooses which inputs to execute on the
  SUT. 
\item \textsf{Test Oracle} evaluates the output
  produced by the SUT and checks if it satisfies the
  specifications.
\end{itemize}
\textsf{Testing Environment} is responsible for
orchestrating the interaction between the components. It queries \textsf{Input Generator} for new inputs to test and it executes them on the SUT. 
\textsf{Testing Environment} collects the output and passes it to \textsf{Test Oracle} for evaluation. 
If the test is complete, \textsf{Testing Environment} stores the final verdict and
resets the environment to start a new test. 
Moreover, the \textsf{Test Oracle}  provides to the \textsf{Input Generator} the set of possible states in which the automaton can currently be, given the executed trace.
In the following, we present each step of our implementation in more detail.

\subsection{Requirements and Automata Processing}
\label{sec:reqs-processing}

The input of the test generator algorithm is a set $R = \{\phi_1,
\dots, \phi_n\}$ of LTL formulas along with the list of
input and output variables.  The parser reads the input formulas as a
conjunction $\Phi = \phi_1 \wedge \dots \wedge \phi_n$ to build the
corresponding automaton. We rely on \spot\cite{spot} to perform
the construction of the B\"uchi automaton represented as a
directed graph. Before test generation starts, we 
preprocess the automaton by expanding the edges
where \spot{} groups different equivalent assignments
to move from one state another, to obtain exactly one assignment for
each edge. During preprocessing, variables are omitted if they are not
relevant for a particular transition, e.g., if the transition is
enabled independently from their 
value. In such cases, we set the input variables to false by default,
while we leave the outputs unchanged. This is because we want to have
a fully defined and deterministic input, but we do not want to impose
additional constraints that are not specified by the requirements on
the outputs. Other choices are possible; for example, one could set
the undefined inputs randomly or could copy the value of such
variables from previous assignments, if any.

\subsection{Test Oracle}
\label{sec:oracle}

The aim of the test oracle is to decide if a trace $\tau$,
composed of input and output variables, is correct with respect to the
given LTL specification $\Phi$. A more permissive check, often
considered for runtime monitoring, consists in verifying that $\tau$ is a
valid prefix of the language $\words(\Phi)$. This can be done by
checking that there exists a run induced by $\tau$ on the automaton 
$A_{\Phi}$, or, equivalently, using monitors.
This kind of check is useful to identify violations of safety
properties, but it is ineffective for liveness ones, even for the
the co-safety subclass. For example, we cannnot detect violations of 
the formula $\phi = \F a$ with a monitor, because every prefix is
valid as long as the proposition $a$ becomes true eventually.  In
order to solve this issue, a number of different LTL semantics for
finite traces have been proposed, such as $FLTL$\cite{manna2012}, 
$LTL^{\mp}$\cite{eisner2003reasoning},
$LTL_3$\cite{bauer2006monitoring} and
$LTL$-$RV$\cite{bauer2010comparing}. In~\cite{bartocci2018counting}
the authors propose a \emph{counting semantics} making predictions
based on the number of steps necessary to witness the satisfaction or
violation of a formula.  Evaluations under such semantics can range
from a 2-valued verdict -- namely \textit{True} ($\top$) or
\textit{False} ($\bot$) -- to a 5-value one; \textit{True}  
($\top$), \textit{Presumably True} ($\top_P$), \textit{Inconclusive}
($?$), \textit{Presumably False} ($\bot_P$) and \textit{False}
($\bot$). The choice of the semantics defines the specific kind of
conformance to the specification adopted and implemented by the test
oracle. In the following, we rely on the FLTL 
semantics, formalized below in Definition \ref{def:fltl} ---
for a discussion of different semantics, we refer the reader 
to~\cite{bauer2010comparing}.
\begin{definition}
	\label{def:fltl}
	Given a finite word (or trace) $\tau$ of length $n$ and an
        FLTL formula $\phi$, $\tau(=\tau, 0)$ satisfies $\phi$,
        denoted as $\tau \models \phi$, under the following conditions
        (s.t. $0 \leq i < n$):
	
	  \begin{tabular}{l l}
	  	 $\tau, i \models p \in AP$ & iff $a \in \tau[i]$ \\
	  	 $\tau, i \models \neg\phi$ & iff $ \tau, i  \not\models \phi$ \\
	  	 $\tau, i \models \phi_1 \wedge \phi_2$ & iff $\tau, i \models \phi_1$ and $\tau, i \models \phi_2$ \\
	  	 $\tau, i\models \X \phi$ & iff $ (i+1 < n)$ and $\tau, i+1 \models \phi$ \\
	  	 $\tau, i \models \mathcal{N} \phi$ & iff $ (i+1 \geq n)$ or $\tau, i+1 \models \phi$ \\
	  	 $\tau, i \models \phi_1 \U\phi_2$ & iff $ \exists i \leq j < n.(\tau, j \models \phi_2 \wedge \forall i \leq m < j.(\tau, m \models \phi_1))$ \\
	  	 $\tau, i \models \F \phi$ & iff $\exists i \leq j < n.(\tau, j \models \phi)$ \\
	  	 $\tau, i \models \G \phi$ & iff $\forall i \leq j < n.(\tau, j \models \phi)$ \\
	  \end{tabular}
\end{definition}
Regarding the boolean operators, FLTL semantics coincides with the
standard LTL semantics on infinite words. However, with temporal
operators, such as $\X$ and $\U$, there is a difference concerning the
maximum length of the word. In particular, the semantics distinguishes
between a strong next operator $\X$, which require a next time step to
exists, and a weak version $\mathcal{N}$, which it is always satisfied
at the last step of a trace. In our requirements, however, we only
make use of the strong variant. 
In our approach, the FLTL oracle is implemented on an automaton and traces are checked directly on the generated  \NBA{}. We posit that every
trace $\tau$ ending in an acceptance state $q^*$ of the 
Automata $A_\Phi$, also satisfies the formula $\Phi$ from which
the automaton is built.



\subsection{Input Generator}

The main idea behind the generation of input sequences for testing the
SUT consists in exploring different paths of the automaton $A_\Phi$
that represents the specification. Given a choice of $(i)$ an
exploration strategy to prioritize paths and $(ii)$ a termination
condition to end the search, we obtain our algorithm Guided Depth First
Search (GDFS) presented in~\ref{alg:GDFS}. As the name suggests, it is
a variant of the classical depth-first search algorithm on directed graphs.

\begin{algorithm}[h]
	\caption{Guided Depth First Search}	
	\label{alg:GDFS}
	\begin{algorithmic}[1]
		\Function{GDFS}{$A_\Phi, k_{min}, k_{max}, oracle, env$}
		\State $visitCounter \gets $ \Call{emptyMap}{ }
		
		\For{$e \in A_\Phi.$\Call{outgoingEdges}{$A_\Phi.initState$}}
			\State $visitCounter[e] \gets 0$
		\EndFor
		
		\While{$\exists e \in visitCounter.(visitCounter[e] == 0)$}
			\State $\tau \gets \{\}$
			\State $s_{c} \gets A_\Phi.initState$
                        \State  $env.$\Call{reset}{ }
			
			\While{$oracle.$\Call{validPrefix}{$\tau$} $\wedge$ $|\tau| < k_{max}$}
			
			\For{$e \in A_\Phi.$\Call{outgoingEdges}{$s_{c}$} $\wedge$  $e \notin visitCounter$}
					\State $visitCounter[e] \gets 0$
			\EndFor
			
			\State $e \gets$ \Call{selectNextEdge}{$A_\Phi, s_{c}, visitCounter}$
			\State $i \gets$  \Call{getInput}{$e$}
			
			\For{$e \in A_\Phi.$\Call{outgoingEdges}{$s_{c}$} $ \wedge$ \Call{getInput}{$e$} $ == i$}
					\State $visitCounter[e] \gets visitCounter[e] + 1$
			\EndFor
			
			\State $o \gets env.$\Call{performAction}{$i$}
			\State $s_{c} \gets$ \Call{getSuccessor}{$A_\Phi, s_{c}, i \cup o$}
			\State $\tau.$\Call{append}{$i \cup o$}
			
			\If{$|\tau| \geq k_{min}$ $\wedge$ $s_{c} \in A_\Phi.acceptanceStates$}
				\State \textbf{break}
			\EndIf
			
			\EndWhile
			
			\State $res \gets oracle.$\Call{evaluate}{$\tau$}
			\State $env.$\Call{addTest}{$\tau, res$}
			
		\EndWhile

		\EndFunction
	\end{algorithmic}
\end{algorithm}

The algorithm takes as input the automaton $A_\Phi$, the interval
$k_{min}$ and $k_{max}$, i.e., the minimum and the maximum length of each trace,
the \textit{oracle} object and the environment \textit{env} object.
The algorithm starts with the initialization of the
\textit{visitCounter} map, that counts how many times
an edge has been explored (lines 2-5). Notice that only the
outgoing edges from the initial state are initialized, while the other
ones are incrementally added during the exploration (lines 11 -
13). The algorithm terminates when all the edges in
\textit{visitCounter} have been visited at least once.  At the
beginning of each test, the trace $\tau$ is initialized to an empty
word and the current state $s_{c}$ is initialized to the initial state
of the automaton (lines 7-8). Then the enviroment is reset to start at
the initial state (line 9).  The test is computed by iteratively
choosing an edge (line 14), extracting the input on its label (line
15), executing it on the SUT by means of the \textit{env} object (line
19) and using the output to choose the successor state, if any, and to
build the trace $\tau$ (lines 20 - 21).  The function
\texttt{selectNextEdge} chooses the next state to execute by selecting
the edge with less visits so far. In case of multiple edges
with the same score, it sorts them with an heuristics that takes into
account the distance from the nearest acceptance state and the degree
of the target state. Moreover, the \textit{visitCounter} is updated
after each choice (lines 16 - 18) by increasing the counter of all
edges leaving $s_{c}$ that present the input $i$. This is a small
optimization to reduce the number of steps necessary to terminate,
because many edges could produce the same input but expect different
accepted outputs. From an input point of view, these edges are
equivalent, but only one of them will be traversed, depending on the
produced output. Termination of a test occurs exactly when
one of the following three cases is true:
$(i)$ $\tau$ is no more a valid prefix of $\mathcal{L}(A_\Phi)$ and
  therefore the test failed;
$(ii)$ the length $\tau$ reached the maximum length $k_{max}$;
$(iii)$ the length of $\tau$ is greater than $k_{min}$ and the
  exploration reached an acceptance state.
At the end of each test, the oracle gives its final verdict and the
result is stored in the $env$ object (lines 26 - 27).

\section{Experimental Analysis}
\label{sec:exp}
We present the results of three experiments\footnote{All benchmarks are available at \url{https://gitlab.sagelab.it/sage/benchmarks-tests}} involving the framework
previously introduced. In the first one, we aim to assess the quality
of the  generated test suite involving a set of benchmarks borrowed by
the LTL Track of the Reactive Synthesis Competition 
2018\footnote{\url{http://www.syntcomp.org/}} (SYNTCOMP 2018). The
second experiment aims to compare the effectiveness of our approach
with respect to model-based strategies; in order to do
that, we consider the use case of an Adaptive Cruise
Control System made available in~\cite{aniculaesei2018automated} and
we compare our algorithm with state-of-the-art model-based approaches
when it comes to spotting erroneous mutants. Finally, our last experiment
aims to evaluate the scalability of our approach in a real world use
case. So, we consider a set of
requirements from the design of an embedded controller for a robotic
manipulator used in the context of the EU project CERBERO\footnote{\url{http://cerbero-h2020.eu}}~\cite{masin2017cross,palumbo2019cerbero}. 
The experiments described in the following ran on a workstation
equipped with an Intel Xeon E31245 @ 3.30GHz CPU and 32GB RAM
running Lubuntu 18.10 64bits. For 
all the experiments, we granted a time limit of 600
CPU seconds (10 minutes) and a memory limit of 30GBs.

\subsection{Syntcomp Benchmarks}
The set of benchmarks we consider is the one provided for the LTL
Track of the Reactive Synthesis Competition 2018.  We first translate
the TLSF~\cite{tlsf} specifications into equivalent LTL ones accepted
by our tool.  Note that we do not use \texttt{SyFCo}, a tool for
manipulating and transforming TLSF specifications in other existing
specification formats for synthesis, because we handle ASSUME formulae
in a different way.  In particular, \texttt{SyFCo} would translate
ASSUME formulae are as a precondition (left-hand side of an
implication) and the ASSERT and GUARANTEE formulae aspostconditions
(right-hand side of an implication).  Therefore, if an ASSUME formula
is violated, the system is not required to satisfy the given
requirements. This behavior would lead to many useless tests, because
whenever an assumption is falsified during the test execution, the
specification would be trivially satisfied and no constraint would be
enforced on the output. In order to solve this problem, we require the
ASSUME part to be satisfied together with the ASSERT and GUARANTEE
part, i.e., we replace implication with conjuction.  We refer the
reader to \cite{tlsf} for more details on the standard translation
from TLSF to LTL.
We exclude benchmarks whose output assignments appear in the
ASSUME part of the specification. This is because, as explained
before, we require the assumptions to hold during the execution of the
test, but assumptions containing outputs can always be falsified, thus
failing the test. 
We sysntesize Mealy machines for the specifications with
\texttt{Strix}~\cite{strix}, the winner of the SYNTCOMP 2018 
competition, and we exclude benchmarks for which \texttt{Strix} times
out in 600 CPU seconds. For each synthesized Mealy machine, we
compute 100 mutants randomly applying one of the following rules:
\begin{itemize}
	\item change the target state of a random transition to a
	different one;
        \item flip the output value of a variable on a
	random transition, namely setting it to \textit{false} if it
	was \textit{true} and vice-versa.
\end{itemize}
We apply only one mutation per mutant because the synthesized models
are  usually small in size and one variation is often enough to expose
a violation of the specification. However, some of the resulting 
mutants may still be correct with respect to the corresponding
specification.
At the end of this process we have 128 different benchmarks,
each of those with 100 mutants.  In the experiment, we compare the
results obtained with 5 different algorithms.  GDFS-1, GDFS-3 and
GDFS-5 are the algorithm described in Section \ref{sec:algo} with
$k_{min}$ set to 1, 3 and 5, respectively.  For comparison purpose, we
also re-implemented, -- and generalized to fit our framework -- the
algorithm presented in \cite{arcaini2013online}.  Briefly, the
algorithm traverses the monitor automaton of the specification during
the test execution, and stops when a coverage criteria is fulfilled. A
test is concluded either when an objective is reached or when the
maximum length $k_{max}$ of the trace is
reached. In \cite{arcaini2013online} two strategies are proposed,
namely Random Walk (RW) and Guided Walk (GW) and we implemented and
tested both of them. As for the coverage criteria, we implemented what
they call \textit{Atomic Proposition Coverage} (APC), i.e., each
atomic proposition on each transition of the monitor must be covered.
For each algorithm we set $k_{max}$ equal to 100 and we stop the
execution as soon as a test fails and the mutant is killed. Notice
that 600 CPU seconds are alloted to each benchmark, including automata
processing and evaluation of all mutants.  

\begin{figure}[t!]
  \begin{tabular}{c@{}c}
    \begin{minipage}{0.5\textwidth}
      \scalebox{0.42}{\includegraphics{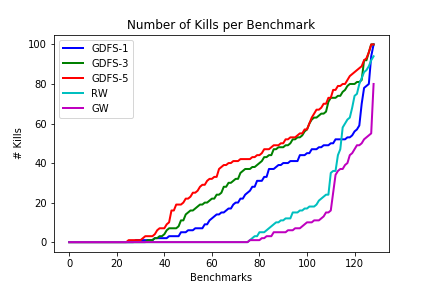}}
    \end{minipage}
    \begin{minipage}{0.5\textwidth}
      \scalebox{0.42}{\includegraphics{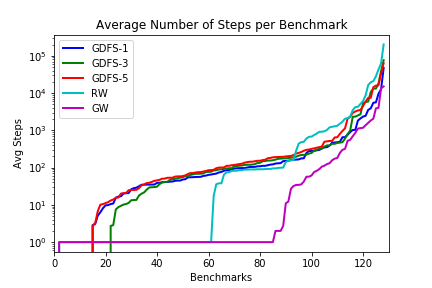}}
    \end{minipage}
  \end{tabular}
  \caption {Total amount of mutants killed (left) and
    average number of steps (right) computed  
    by the considered algorithms in the set of SYNTCOMP 2018 benchmarks.}
  \label{fig:kills-plot}
\end{figure}

Figure \ref{fig:kills-plot} (left) shows the number of mutants killed
per benchmark by each algorithm, ranging from 0 to 100. 
Figure \ref{fig:kills-plot} (right) shows the average number of steps
executed, namely the sum of the length of each test, averaged over the
mutants.  In both charts, the abscissa represents the number of
benchmarks, while the ordinate shows the number of mutants killed (left)
and the number of steps executed (right).  Notice that, since the
results of RW and GW can vary due to non-deterministic behaviors, we
execute the test 3 times and we report the median value as reference
for these two algorithms.  The results reveal that GDFS-5 clearly
outperform all the other algorithms in terms of total amount of
mutants killed, and that the number of executed steps is only slightly
higher than GDFS-1 and GDFS-3. However, only for two benchmarks all
the 100 mutants have been killed. Moreover, in 25 cases it did not
kill any mutant, 15 of which due to timeouts.  Regarding RW and GW,
they both revealed totally ineffective for 73 of the 129 benchmarks,
although only 2 timeouts occurred. However, looking at
Figure \ref{fig:kills-plot} (right) we notice that in 59 of these
benchmarks, the two algorithms did not perform any testing at all. This
phenomenon is due to the nature of the benchmarks involved, where the
specification only contains liveness properties and the monitor is a
single state automaton accepting all prefixes.

\subsection{Adaptive Cruise Control}
In our second experiment we consider the Adaptive Cruise Control (ACC)
prototype implemented in~\cite{aniculaesei2018automated}.  The ACC
system adjusts the current velocity of the vehicle
towards a target cruise velocity defined by driver. If the
vehicle gets too close to the forward vehicle, the ACC system must
adjust the current distance between the two and maintain a certain
safety distance. Additionally, the driver can intervene by: (1)
activating the system via an ACC button; (2) deactivating the system
via the ACC button; and (3) deactivating the system by braking or
accelerating the car.  The authors of~\cite{aniculaesei2018automated} also
generated test cases from LTL requirements using three different
requirements coverage criteria: requirements coverage (RC), antecedent
coverage (AC), and unique first cause coverage (UFC). Tests are
generated with a model-based generation strategy: trap-properties are
built from requirements, and a counterexample is produced with a model
checker. The algorithms are evaluated with 524 mutants of the correct
implementation.

The goal of the experiment here described is to compare the
performance of our algorithm with respect to model-based techniques
that make explicit use of a model to generate test cases.  We 
modified slightly the set of requirements, reducing numerical
comparisons and enums (available in the NuSMV~\cite{cimatti2002}
models used in \cite{aniculaesei2018automated}) to boolean
variables. This is a mere syntactic variation to represents LTL
formulae in the default syntax as described in Section
\ref{sec:LTLdef}.  
The resulting specification is composed of 12 requirements, 6 input
and 10 output variables. The results are depicted in Table
\ref{tab:acc-results}. In order to ease the comparison with the
model-based approach, we also report the results from
\cite{aniculaesei2018automated}. 

\begin{table}[t!]
	\caption{Experimental results on the ACC use case.}
	\label{tab:acc-results}
	\centering
        \scriptsize
	\begin{tabular}{|l|r|r|r||r|r|r|}
		\hline
		& \textbf{RC} & \textbf{AC} & \textbf{UFC} & \textbf{GDFS-1} & \textbf{GDFS-3} &  \textbf{GDFS-5} \\
		\hline
		Number of Test Cases & 6 & 7 & 18 & 26 & 4912 & 2597 \\
		\hline
		Branch Coverage (\%) & 78.3 & 78.3 & 86.7 & 45.0 & 70.0 & 71.7 \\
		\hline
		Number of Killed Mutants & 488 & 488 & 488 & 414 & 480 & 480 \\
		\hline
		Killed Mutants (\%) & 93.1 & 93.1 & 93.1 & 79.0 & 91.6 & 91.6 \\
		\hline
	\end{tabular}
	\vspace{-3mm}
\end{table}

The results show that the GDFS algorithm performances are comparable to
the model-based algorithms, with a difference of only 8 mutants
($1.5\%$ of the total) for $k_{min}$ equal 3 or 5, at the expense of
many more tests. Notice however that the test generation and execution
is still quite small; it takes about 1 second to run GDFS-1, 11
seconds for GDFS-3 and 5 seconds for GDFS-5. Moreover, the whole test
suite is executed only if all tests succeed, but if a failure is
detected it can terminate much earlier. In the case of GDFS-5, for
example, the average number of tests executed per mutant is 329, much
lower than the test suite size (2597).  However, despite the large
test suite, GDFS reaches a lower branch coverage than the model-based 
counterparts, stopping at $71.7\%$.  
Also notice that, in this context, with all requirements being safety
properties, the RW algorithm described in the previous experiment
performs well, achieving similar results to GDFS-5 (although with some
variation due to randomness). 
These results show that the black-box testing with the framework
presented in Section \ref{sec:algo} can be almost as effective as
model-based techniques, where more manual work is required to model
the system. 
A final remark on the $k_{min}$ and $k_{max}$ parameters of the GDFS
algorithm is in order. As shown in Table \ref{tab:acc-results},
$k_{min}$ plays an important role in the test suite size and
performance. In our experience, the longer the test, the more the
automaton is covered and the less transitions close to the initial
state are repeated. Similarly, also $k_{max}$ can influence a test
suite size and performance: an excessively small value could lead to
some false positive tests, while an excessively large value could 
produce unnecessarily long tests before declaring them failed.  
However, the generated test suite depends not only on the algorithm
and the specification, but also on the SUT behavior. The optimal
values of such parameters is context dependent, and may require some
fine tuning.

\subsection{Robotic Manipulator}

Our last experiment considers a set of requirements from the design of
an embedded controller for a robotic manipulator. The controller
should direct a properly initialized robotic arm --- and related
vision system --- to look for an object placed in a given position and
move to such position in order to grab the object; once grabbed, the
object has to be moved and released into the bucket without touching
it. The robot must stop also in the case of an unintended collision
with other objects or with the robot itself --- collisions can be
detected using torque estimation from current sensors placed in the
joints. Finally, if a general alarm is detected, e.g., by the
interaction with a human supervisor, the robot must stop as soon as
possible.  The manipulator is a 4 degrees-of-freedom Trossen Robotics
WidowX
arm\footnote{ \url{http://www.trossenrobotics.com/widowxrobotarm}.}
equipped with a gripper.  The design of the embedded controller is
part of the activities related to the ``Self-Healing System for
Planetary Exploration'' use case in the context of the EU project
CERBERO.  In this case the specification is composed of 31
requirements, 3 inputs and 11 outputs.  The SUT is implemented as an
smv model.  With GDFS-5 ($k_{min}$ = 5 and $k_{max}$ = 30), we
obtain 1441 tests and a total of 12867 steps executed in 1171
seconds.  At each step, NuSMV~\cite{cimatti2002} is called in order to
determine the evolution of the system.
Then, we manually inject faults by removing some constraints in the
guards (forcing the system to evolve from one state to another) or by
modifying value assignments of some variables. At the end, we obtain
10 different NuSMV faulty models. We show the results of this analysis
in Table \ref{tab:inj-results}. First, we report that a failed test
has been detected in all considered cases. Looking at the Table, we
can observe that, for each bugged system, a small number of tests is
necessary to discover the failure. Therefore, in most cases, it is not
necessary to perform a complete exploration of the automaton and an
early stopping strategy can save substantial time when debugging an
application.  

\begin{table}[!t]
	\caption{Fault-Injection results on the robotic manipulator use case.}
	\label{tab:inj-results}
	\centering
	\begin{tabular}{|l|r|r|r|}
		\hline
		\textbf{\# Injection} & \textbf{\# Tests} & \textbf{\# Steps} & \textbf{Time(s)} \\
		\hline
		1 & 1 & 2  & 7.64 \\ \hline
		2 & 2 & 14 & 8.61 \\ \hline
		3 & 2 & 14 & 8.74 \\ \hline
		4 & 1 & 2  & 7.75 \\ \hline
		5 & 1 & 7  & 8.15 \\ \hline
		6 & 4 & 25 & 8.61 \\ \hline
		7 & 56 & 502 & 25.23 \\ \hline
		8 & 1 & 3 & 8.15 \\ \hline
		9 & 1 & 6 & 7.84 \\ \hline
		10 & 2 & 10 & 8.17\\ \hline
	\end{tabular}
\end{table}

\section{Conclusions}
\label{sec:concl}
 
In this paper, we have described a new approach to conformance 
testing of black-box reactive systems. 
We evaluated our approach across three different experimental
settings. In the first setting we synthesized a set of benchmarks taken
from the SYNTCOMP 2018 competition and we showed that our approach
is better at finding mutants than (a generalization of) two
different algorithms presented in~\cite{arcaini2013online}. In the
second setting, we showed that our approach compares favorably
with state-of-the-art model-based techniques.
Finally, in the third setting we tested a controller for a robotic
manipulator modeled in \smv{} and we showed that our approach
is able to find some manually injected faults.
As future work, we plan to $(i)$ extend the framework with more test
oracles and exploration strategies and $(ii)$ increase the input language
expressiveness with the addition of numerical constraints. 
The implementation of our approach is freely available in the
SpecPro\footnote{\url{https://gitlab.sagelab.it/sage/SpecPro}} Java
library. 


\paragraph{Acknowledgments} The research of Luca Pulina and Simone Vuotto is part of the FitOptiVis project funded by the ECSEL Joint Undertaking under grant number H2020-ECSEL-2017-2-783162. The research of Luca Pulina has been also partially funded by the ECSEL JU Project COMP4DRONES and the Sardinian Regional Projects PROSSIMO (POR FESR Sardegna 2014/20-ASSE I).

\newpage

\bibliographystyle{abbrv} 




\end{document}